\documentclass{iopjournal}

\usepackage{graphicx}
\usepackage{amssymb}
\usepackage{amsmath}
\usepackage{url}
\usepackage{placeins}
\usepackage{fancyhdr}
\usepackage{ragged2e}

\begin{document}
\justifying

\articletype{Paper}

\title{BORAY-3D: A ray tracing code for three-dimensional magnetized plasma configurations}

\author{Yuxuan Wang$^{1,2}$ and Huasheng Xie$^{1,*}$}

\affil{$^1$Beijing VeloAlpha Technology Co., Ltd., Beijing 100080, China}

\affil{$^2$National University of Defense Technology, Changsha 710049, China}


\email{huashengxie@gmail.com}


\begin{abstract}
\justifying
Ray tracing codes are useful tools for studying electromagnetic wave propagation and absorption using the geometrical-optics approximation. Existing codes commonly provide either broad radio-frequency coverage in axisymmetric equilibria or three-dimensional capability specialized for electron-cyclotron (EC) applications. BORAY-3D integrates three desirable features in a single version. First, it has a broad frequency range of validity regime from ion-cyclotron, helicon and lower-hybrid waves to EC waves and emission. Second, it provides an unified treatment of arbitrary two- and three-dimensional magnetic-plasma configurations, including both closed and open field-line regions. Third, it incorporates fully relativistic Maxwellian EC absorption. The code extends the axisymmetric BORAY formulation by solving the ray equations in cylindrical coordinates $(r,\phi,z)$ while allowing the toroidal mode number $n_\phi$ to vary. Magnetic-field, density and temperature data are directly described in $(r,\phi,z)$ coordinates without the restriction of flux functions, so that numerical equilibria and analytic field models can be handled in the same form. The non-relativistic hot-plasma model inherited from BORAY is used for lower-hybrid, ion-cyclotron and helicon absorption, whereas the relativistic model is coupled to reciprocal radiative transfer for electron cyclotron emission (ECE). Practical applications include 13.56 MHz helicon and 50 MHz fast waves, a 3.7 GHz lower-hybrid wave, and 115--220 GHz EC emission. BORAY-3D has been systematically benchmarked against GENRAY for tokamak toroidal-field ripple, Raytrax and TRAVIS for W7-X, as well as public HSX heating and W7-X ECE results.
\end{abstract}

\section{Introduction}
\label{sec:intro}

RF waves are widely used for plasma heating and current drive in magnetic-confinement devices. The corresponding frequency spectrum can vary from megahertz (MHz) to gigahertz (GHz), based on specific heating processes, including ion-cyclotron radio frequency heating (ICRH), lower hybrid wave (LHW), and electron-cyclotron frequency heating (ECRH). Electron cyclotron emission (ECE) at the same high-frequency range is routinely used to measure the electron temperature of ECRH \cite{Bornatici1983,Hartfuss1997}. When the wavelength is much shorter than the equilibrium scale length, wave propagation and absorption can be described by the geometric optical approximation and the corresponding ray tracing model \cite{Stix1992}. A unified numerical framework incorporating above RF waves is useful because comprehensive physics capabilities can be applied on the same footing for different plasma equilibria, and the appropriate absorption model is selected for each wave regime.

The ray tracing code BORAY \cite{Xie2022} has been developed in cylindrical coordinates $(r,\phi,z)$ for arbitrary axisymmetric plasmas with closed or open magnetic field lines. Its fluid and kinetic dispersion-relation solvers were developed based on the theoretical frameworks PDRF \cite{Xie2014PDRF}, PDRK \cite{Xie2016PDRK} and BO \cite{Xie2019BO}. A striking feature of BORAY is one-solve-all RF waves with the same physics model equations and numerical implementation. It has been benchmarked for ECRH, lower-hybrid waves (LHW), helicon waves and ICRH in tokamak, spherical-tokamak, field-reversed and mirror configurations. Because density and temperature are expressed directly in cylindrical coordinates without the constraint of magnetic-flux functions, closed and open field-line regions can be incorporated in the same simulation. However, the original implementation was restricted to two-dimensional axisymmetric profiles and omitted relativistic absorption \cite{Xie2022}. It therefore cannot describe stellarators such as W7-X \cite{Beidler2021} or the toroidal-field ripple produced by a finite number of tokamak coils, which can change the ray trajectory and power deposition position \cite{Peysson2012}.

Several early work have been carried out on the topic of three-dimensional ray tracing modeling. TRAVIS \cite{Marushchenko2014} can model ECRH and current drive and ECE in stellarator equilibria using VMEC flux coordinates \cite{Hirshman1983}. The open-source code Raytrax \cite{Raytrax} is also developed to study the ECRH propagation and absorption in stellarator equilibria. Several engineering codes have been developed for specific machines with 3D equilibria. LHDGauss is a multi-ray code for EC beam propagation and deposition in LHD, which incorporates experimentally reconstructed three-dimensional equilibria and oblique-propagation absorption \cite{Tsujimura2015LHDGauss,Yanai2021LHDGauss}. TRECE performs three-dimensional Hamiltonian ray tracing for EC heating and current drive in TJ-II \cite{Tribaldos1998TRECE}, and TRUBA provides both ray- and beam-tracing descriptions, including electron-Bernstein-wave propagation and resonance mode conversion in the same device \cite{Castejon2004TRUBA}. Beyond conventional ray or multi-ray descriptions, the quasioptical code PARADE evolves a finite-width wave beam with diffraction and resonance mode conversion and has been applied to EC propagation and absorption in LHD \cite{Yanagihara2021PARADE}. GENRAY includes analytic toroidal-ripple models in tokamak. Above studies establish accurate three-dimensional ECRH modeling, but do not provide the same capability for other RH waves, geometry-independent cylindrical input and relativistic EC absorption pursued here.

In this work, we develop a new code for the applications at different frequency regimes and geometries. First, BORAY-3D retains the broad-frequency capability of BORAY, covering low-frequency ICW and helicon waves, intermediate-frequency LHW and high-frequency ECW/ECE with the corresponding kinetic absorption treatments. Second, two-dimensional axisymmetric and unified three-dimensional configurations are represented in a cylindrical-coordinate formulation. The equilibrium parameters, e.g. magnetic field, density and temperature, are expressed using $(r,\phi,z)$ rather than the flux functions, which can treat both closed and open field-line regions. Third, the non-relativistic limitation of the original BORAY code is removed for EC applications by implementing fully relativistic Maxwellian absorption and reciprocal ECE radiative transfer. The applications of BORAY-3D cover various RF waves from 13.56 MHz to 220 GHz and tokamak-ripple, stellarator and helical configurations. The remainder of this paper is organized as follows. The physics models for ray propagation and absorption are described in Sec. \ref{sec:equations}. The benchmarks and practical applications are performed in Sec. \ref{sec:benchmarks}. The conclusion is given in Sec. \ref{sec:summary}.

\section{Physics model}
\label{sec:equations}

\subsection{Ray tracing equations in three-dimensional cylindrical coordinates}
\label{sec:rayeq}

Following the framework of BORAY code \cite{Xie2022}, we use cylindrical coordinates $(r,\phi,z)$ with wave vector variables $(k_r,n_\phi=rk_\phi,k_z)$, where $n_\phi$ is the Fourier coordinate conjugate of $\phi$. The geometric optical equations are
\begin{equation}
\frac{dr}{d\tau}=\frac{\partial D}{\partial k_r},\quad
\frac{d\phi}{d\tau}=\frac{\partial D}{\partial n_\phi},\quad
\frac{dz}{d\tau}=\frac{\partial D}{\partial k_z},
\label{eq:ray1}
\end{equation}
\begin{equation}
\frac{dk_r}{d\tau}=-\frac{\partial D}{\partial r},\quad
\frac{dn_\phi}{d\tau}=-\frac{\partial D}{\partial \phi},\quad
\frac{dk_z}{d\tau}=-\frac{\partial D}{\partial z},
\label{eq:ray2}
\end{equation}
with
\begin{equation}
\frac{dt}{d\tau}=-\frac{\partial D}{\partial \omega},
\end{equation}
where $D(\omega,k_\parallel^2,k_\perp^2,r,\phi,z)=0$ determines the dispersion relation, $k_\parallel=\mathbf{k}\cdot\mathbf{b}=\frac{1}{B}(k_rB_r+k_zB_z+\frac{n_\phi}{r}B_\phi)$, $k_\perp^2=k^2-k_\parallel^2$ and $k^2=k_r^2+k_z^2+n_\phi^2/r^2$. In BORAY the axisymmetry assumption $\partial D/\partial\phi=0$ makes $n_\phi$ a conserved quantity. In BORAY-3D we have $B=B(r,\phi,z)$ and $n_{s0}=n_{s0}(r,\phi,z)$, where the full $\phi$ dependence is retained. Defining
\begin{equation}
D_\parallel\equiv\frac{\partial D}{\partial k_\parallel^2},\quad
D_\perp\equiv\frac{\partial D}{\partial k_\perp^2},\quad
A\equiv D_\parallel-D_\perp,
\end{equation}
the explicit form of Eqs. (\ref{eq:ray1}) and (\ref{eq:ray2}) is
\begin{eqnarray}
\frac{dr}{d\tau}&=&2Ak_\parallel\frac{B_r}{B}+2D_\perp k_r,\\
\frac{d\phi}{d\tau}&=&2Ak_\parallel\frac{B_\phi}{rB}+2D_\perp\frac{n_\phi}{r^2},\\
\frac{dz}{d\tau}&=&2Ak_\parallel\frac{B_z}{B}+2D_\perp k_z,
\end{eqnarray}
and
\begin{eqnarray}
\frac{dk_r}{d\tau}&=&-\frac{\partial D}{\partial r}\Big|_{k_\parallel^2,k_\perp^2}
-2Ak_\parallel\frac{\partial k_\parallel}{\partial r}
+2D_\perp\frac{n_\phi^2}{r^3},\\
\frac{dn_\phi}{d\tau}&=&-\frac{\partial D}{\partial \phi}\Big|_{k_\parallel^2,k_\perp^2}
-2Ak_\parallel\frac{\partial k_\parallel}{\partial \phi},
\label{eq:dnphi}\\
\frac{dk_z}{d\tau}&=&-\frac{\partial D}{\partial z}\Big|_{k_\parallel^2,k_\perp^2}
-2Ak_\parallel\frac{\partial k_\parallel}{\partial z},
\end{eqnarray}
where
\begin{eqnarray}
\frac{\partial k_\parallel}{\partial r}&=&
-\frac{k_\parallel}{B}\frac{\partial B}{\partial r}
+\frac{1}{B}\Big(k_r\frac{\partial B_r}{\partial r}
+k_z\frac{\partial B_z}{\partial r}
+\frac{n_\phi}{r}\frac{\partial B_\phi}{\partial r}
-\frac{B_\phi n_\phi}{r^2}\Big),\\
\frac{\partial k_\parallel}{\partial \phi}&=&
-\frac{k_\parallel}{B}\frac{\partial B}{\partial \phi}
+\frac{1}{B}\Big(k_r\frac{\partial B_r}{\partial \phi}
+k_z\frac{\partial B_z}{\partial \phi}
+\frac{n_\phi}{r}\frac{\partial B_\phi}{\partial \phi}\Big),
\label{eq:dkpardphi}\\
\frac{\partial k_\parallel}{\partial z}&=&
-\frac{k_\parallel}{B}\frac{\partial B}{\partial z}
+\frac{1}{B}\Big(k_r\frac{\partial B_r}{\partial z}
+k_z\frac{\partial B_z}{\partial z}
+\frac{n_\phi}{r}\frac{\partial B_\phi}{\partial z}\Big).
\end{eqnarray}

For wave propagation, the cold plasma dispersion relation is used \cite{Stix1992,Xie2014PDRF},

\begin{eqnarray}
D_{\rm cold}(\omega,k_\parallel^2,k_\perp^2)&=&
\varepsilon_1\frac{k_\perp^4c^4}{\omega^4}
-\Big[(\varepsilon_1+\varepsilon_3)\Big(\varepsilon_1-\frac{k_\parallel^2c^2}{\omega^2}\Big)-\varepsilon_2^2\Big]\frac{k_\perp^2c^2}{\omega^2}\nonumber\\
&&+\,\varepsilon_3\Big[\Big(\varepsilon_1-\frac{k_\parallel^2c^2}{\omega^2}\Big)^2-\varepsilon_2^2\Big]=0,
\label{eq:colddr}
\end{eqnarray}
with
\begin{equation}
\varepsilon_1=1-\sum_s\frac{\omega_{ps}^2}{\omega^2-\omega_{cs}^2},\quad
\varepsilon_2=\sum_s\frac{\omega_{cs}}{\omega}\frac{\omega_{ps}^2}{\omega^2-\omega_{cs}^2},\quad
\varepsilon_3=1-\sum_s\frac{\omega_{ps}^2}{\omega^2},
\end{equation}
where $\omega_{cs}=q_sB/m_s$ and $\omega_{ps}=\sqrt{n_{s0}q_s^2/\epsilon_0m_s}$. The treatment of the cyclotron resonance singularity of Eq. (\ref{eq:colddr}) is identical to BORAY, being reported in Ref. \cite{Xie2022}. The time integration in Eqs. (\ref{eq:ray1}) and (\ref{eq:ray2}) applies a fourth-order Runge-Kutta scheme.

It should be noted that the above mathematical formulation is valid for both axisymmetric and non-axisymmetric, both open and closed field line plasmas, which will be used for various magnetized plasma configurations in this work.

\subsection{Wave absorption}
\label{sec:absorption}

For non-relativistic waves, the absorption is calculated from the Maxwellian hot-plasma dispersion tensor used in BORAY \cite{Yu2026Absorption}. At each position on the cold-plasma ray, the complex $k_\perp$ is obtained from
\begin{equation}
\det\boldsymbol{D}_{\rm hot}(\omega_0,k_\perp,k_\parallel)=0 .
\label{eq:iterationki}
\end{equation}
The real frequency and the signed $k_\parallel$ are fixed, and the hot-plasma root is followed continuously from the real cold-plasma root. The R\"onnmark form of the dielectric tensor \cite{Ronnmark1983} is evaluated using a 12-pole expansion of the plasma dispersion function \cite{Xie2016PDRK,Xie2019BO,Xie2024PDF} and the WHAMP finite-Larmor-radius functions \cite{Ronnmark1982}. The remaining power is
\begin{equation}
\frac{P(s)}{P_0}=\exp\left[-2\int_0^s
\boldsymbol{k}_i\boldsymbol{\cdot}d\boldsymbol{r}\right].
\label{eq:power}
\end{equation}
The power absorbed by each species is calculated from its anti-Hermitian contribution to the dielectric tensor. With $\boldsymbol{\epsilon}_s^A=(\boldsymbol{\epsilon}_s-\boldsymbol{\epsilon}_s^\dagger)/(2i)$, the local fraction is
\begin{equation}
\eta_s=\frac{\widehat{\boldsymbol e}^{\dagger}\boldsymbol{\epsilon}_s^A
\widehat{\boldsymbol e}}
{\widehat{\boldsymbol e}^{\dagger}\sum_a\boldsymbol{\epsilon}_a^A
\widehat{\boldsymbol e}},
\qquad
P_s(s)=\int_0^s\eta_s(s')\left[-\frac{dP}{ds'}\right]ds'.
\label{eq:speciespartition}
\end{equation}

\subsection{Relativistic EC absorption and ECE radiative transfer}
\label{sec:ece}

For electron cyclotron waves, relativistic effects shift and broaden the resonance and are important for weakly absorbed higher harmonics \cite{Bornatici1983}. BORAY-3D uses the fully relativistic Maxwellian absorption integral of Refs. \cite{Albajar2007,Marushchenko2014}. The wave polarization is calculated from the weakly relativistic Krivenski--Orefice tensor \cite{Krivenski1983}. The power flux can be evaluated from either the cold-plasma tensor or the R\"onnmark hot-plasma tensor \cite{Ronnmark1983}, with the latter used by default.

All quantities below are written in the local Stix frame. Its unit vectors are $\widehat{\boldsymbol{z}}=\boldsymbol{B}/B$, $\widehat{\boldsymbol{x}}=\boldsymbol{k}_{\perp}/k_{\perp}$ and $\widehat{\boldsymbol{y}}=\widehat{\boldsymbol{z}}\times\widehat{\boldsymbol{x}}$. The refractive-index vector is consequently $\boldsymbol{N}=(N_{\perp},0,N_{\parallel})^{\mathsf T}=c\boldsymbol{k}/\omega$. Define $\omega_{pe}^{2}=n_e e^2/(\epsilon_0m_e)$, $\Omega_e=-eB/m_e$ and $Y=|\Omega_e|/\omega$. The normalized electron momentum is $\boldsymbol{p}=\boldsymbol{P}/(m_ec)=(p_{\perp},p_{\parallel})$, with Lorentz factor $\gamma=(1+p_{\perp}^{2}+p_{\parallel}^{2})^{1/2}$. The inverse normalized temperature is $\mu=m_ec^2/T_e$, and the Maxwell--J\"uttner distribution is
\begin{equation}
f_M(\boldsymbol{p})=\frac{\mu}{4\pi K_2(\mu)}
\exp(-\mu\gamma),
\label{eq:juttner}
\end{equation}
where $T_e$ is expressed in energy units and $K_2$ is the modified Bessel function of the second kind.

Let $\widehat{\boldsymbol{e}}=(e_x,e_y,e_z)^{\mathsf T}$ denote the unit electric-field polarization, $\widehat{\boldsymbol{e}}^{\dagger}\widehat{\boldsymbol{e}}=1$, and let $\boldsymbol{F}$ be the dimensionless power-flux vector defined in Appendix \ref{app:relativistic}. The absorption coefficient is
\begin{equation}
\alpha_\omega=
\frac{2\pi^2\omega_{pe}^2}{c\,\omega\,|\boldsymbol{F}|}
\sum_{n=1}^{N_h}
\int_{p_{\parallel,n}^{-}}^{p_{\parallel,n}^{+}}
dp_\parallel\,\mu f_M(\boldsymbol{p})\,|\mathcal{C}_n|^2 ,
\label{eq:relabs}
\end{equation}
where $N_h$ is the highest retained cyclotron harmonic and $\boldsymbol F$ is the dimensionless power-flux vector. The harmonic coupling is
\begin{equation}
\mathcal{C}_n=
e_x\frac{nY}{N_{\perp}}J_n(\xi)
-i e_y p_{\perp}J_n'(\xi)
+e_zp_{\parallel}J_n(\xi),
\label{eq:harmoniccoupling}
\end{equation}
where $\xi=N_{\perp}p_{\perp}/Y$, and $J_n$ and $J_n'$ are the Bessel function of the first kind and its derivative. On the resonance curve, $\gamma=nY+N_{\parallel}p_{\parallel}$ and $p_{\perp}^{2}=\gamma^2-1-p_{\parallel}^{2}$. The integration limits follow from $p_{\perp}=0$,
\begin{equation}
p_{\parallel,n}^{\pm}=
\frac{nYN_{\parallel}\pm
\sqrt{(nY)^2+N_{\parallel}^{2}-1}}
{1-N_{\parallel}^{2}}.
\label{eq:resbounds}
\end{equation}
A harmonic contributes when the resonance condition in Eq. (\ref{eq:resbounds}) has real limits. The remaining power is $P(s)/P_0=\exp[-\int_0^s\alpha_\omega(s')ds']$. The Krivenski--Orefice and R\"onnmark tensors, together with the power-flux expression, are given in Appendix \ref{app:relativistic}.

The ECE module computes the radiative temperature spectrum received by an antenna \cite{Bornatici1983,Hartfuss1997}. For each frequency channel, a reciprocal ray is launched from the receiver position along the line of sight and propagated through vacuum to the plasma boundary. Along the ray, the absorption coefficient $\alpha_\omega$ gives the optical depth
\begin{equation}
\tau(s)=\int_0^s\alpha_\omega(s')\,ds'
\label{tau}
\end{equation}
\noindent so that the remaining wave power is
$P(s)/P_0=\exp[-\tau(s)]$. For a Maxwellian plasma, Kirchhoff's law in the Rayleigh--Jeans limit yields the radiative temperature
\begin{equation}
T_{\mathrm{ECE}}(\omega)=\int_0^{\tau_{\mathrm{end}}}T_e(\tau)\,e^{-\tau}\,d\tau.
\label{eq:trad}
\end{equation}
Either the non-relativistic kinetic model or the relativistic model of Eq. (\ref{eq:relabs}) can supply $\alpha_\omega$.

\section{Benchmarks and applications}
\label{sec:benchmarks}

This section verifies the three combined capabilities mentioned above: frequency coverage from megahertz to hundreds of gigahertz, an unified treatment of axisymmetric and fully three-dimensional configurations, and relativistic EC absorption and emission. The benchmarks and applications are summarized in Table \ref{tab:benchmarks} and are compared with other ray tracing codes or published results. All cases use the ray equations in Sec. \ref{sec:rayeq}. The non-relativistic hot-plasma model is used for LHW, ICW and helicon absorption, while the relativistic model in Sec. \ref{sec:ece} is used for ECW absorption and ECE. As in BORAY, ray tracing results for low-frequency waves should be used with caution when the wavelength is not sufficiently shorter than the equilibrium scale length. In such cases, quantitative assessment usually requires comparison with experiment or full-wave calculations \cite{Xie2022}.

\begin{table}[htbp]
\caption{BORAY-3D benchmark and application examples.}
\label{tab:benchmarks}
\centering
\small
\begin{tabular}{llll}
\hline
Case & Wave & Reference & Result\\
\hline
MUSE Helicon & 13.56 MHz &  & Fig. \ref{fig:muse_helicon}\\
LHD ICRF & 50 MHz Fast wave & Saito et al.\ \cite{Saito2010} & Fig. \ref{fig:lhd_icrf}\\
Tokamak TF Ripple & 3.7 GHz LHW & GENRAY & Fig. \ref{fig:ppcf_lh}\\
HSX ECW & 28 GHz X2 & Likin et al.\ \cite{Likin2003} & Fig. \ref{fig:hsx}\\
W7-X ECW & 140 GHz O & Raytrax, TRAVIS & Figs. \ref{fig:w7x}, \ref{fig:d1}\\
W7-X ECE & 115--220 GHz X & Marushchenko et al.\ \cite{Marushchenko2007} & Fig. \ref{fig:ece}\\
\hline
\end{tabular}
\end{table}

\subsection{Tokamak toroidal field ripple}
\label{sec:genray}

We first compare BORAY-3D and GENRAY for lower-hybrid wave in a tokamak with toroidal-field ripple. The circular-tokamak parameters $R_0=3.05$ m, $a=0.95$ m, $B_0=3.2$ T and $I_p=3.5$ MA are taken from Peysson et al. \cite{Peysson2012}. An analytic 18-coil ripple model is used in both codes. To make the ripple effect clearly visible, the ripple amplitude at the boundary is increased to $\delta_{\mathrm{rip}}=6\%$. A 3.7 GHz slow-LH wave is launched at $(\rho,\theta,\phi)=(0.968,15^\circ,0)$ with $(N_\phi,N_\theta)=(-2,0)$. The profiles are $n_e=[0.01+4.99(1-\rho^2)]\times10^{19}\ \mathrm{m}^{-3}$ and $T_e=[0.1+4.9(1-\rho^2)]$ keV for $\rho\leq1$.

The ray trajectories from BORAY-3D and GENRAY are almost identical in both the axisymmetric and ripple cases. Their maximum position differences are 1.0 mm and 7.1 mm, respectively, whereas the displacement caused by ripple reaches 0.72 m. Both codes give a change of the minimum radius from $\rho\simeq0.18$ to $\rho\simeq0.70$ and an increase of the maximum $|N_\parallel|$ from 2.8 to about 4.7. The power absorption also agrees well for the ripple case. The larger difference in the axisymmetric case is mainly due to the different absorption models: GENRAY uses the Bonoli lower-hybrid model, while BORAY-3D uses the non-relativistic Maxwellian model in Sec. \ref{sec:absorption}.

\begin{figure}[htbp]
\centering
\includegraphics{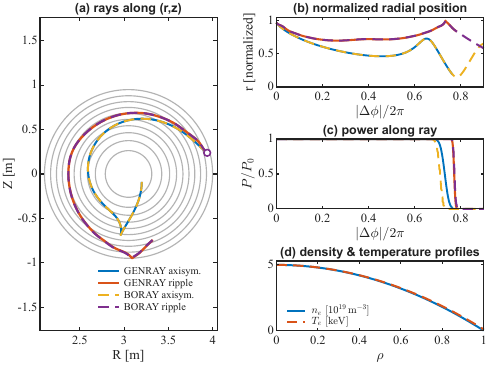}
\caption{Comparison of BORAY-3D and GENRAY for a 3.7 GHz lower-hybrid wave in a tokamak with strong toroidal-field ripple. Solid and dashed curves denote GENRAY and BORAY-3D, respectively. (a) Poloidal ray trajectories, (b) normalized radial position, (c) remaining power and (d) density and electron-temperature profiles. The ray trajectories agree well. The difference in power absorption is due mainly to the different hot-plasma models.}
\label{fig:ppcf_lh}
\end{figure}
\FloatBarrier

\subsection{Stellarator ECW (O and X modes)}
\label{sec:w7x}

In this section, we show the capability of BORAY-3D for stellarator configuration, for which this work are mainly motivated to provide support.

\begin{figure}[htbp]
\centering
\includegraphics{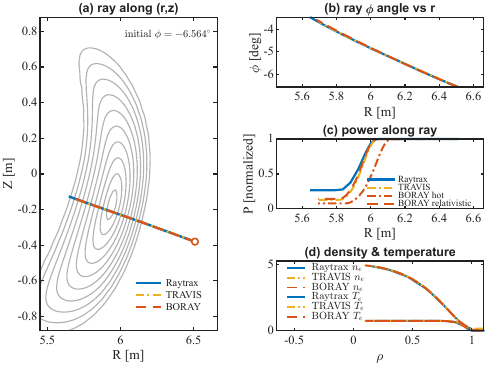}
\caption{Comparison of BORAY-3D, Raytrax and TRAVIS for a W7-X 140 GHz O-mode ray. The trajectories agree within 5 mm. The differences in remaining power are mainly due to the different absorption models.}
\label{fig:w7x}
\end{figure}

Fig. \ref{fig:w7x} compares a W7-X 140 GHz O-mode ray calculated with BORAY-3D, TRAVIS and Raytrax. The public W7-X standard vacuum equilibrium has five field periods and $B_0=2.52$ T. The electron profiles are $n_e(0)=0.75\times10^{20}\ \mathrm{m}^{-3}$ and $T_e(0)=5$ keV. The launch position is $(R,\phi,Z)=(6.509\ \mathrm{m},-6.56^\circ,-0.38\ \mathrm{m})$, with poloidal and toroidal aiming angles $(15.7^\circ,19.7^\circ)$. The three trajectories agree within 5 mm. The final absorbed fractions are 85.7\%, 87.1\% and 73.3\% for BORAY-3D, TRAVIS and Raytrax, respectively. The difference is due mainly to the non-identical kinetic absorption models.

\begin{figure}[htbp]
\centering
\includegraphics{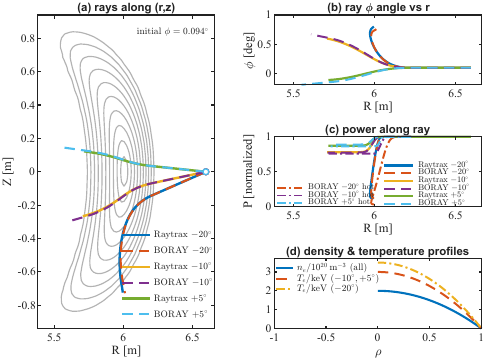}
\caption{Comparison of BORAY-3D and Raytrax for three W7-X Port-A D1 launches with poloidal aiming angles $-20^\circ$, $-10^\circ$ and $+5^\circ$. The ray trajectories and final absorbed fractions agree well for all three angles.}
\label{fig:d1}
\end{figure}

Fig. \ref{fig:d1} compares BORAY-3D and Raytrax for three W7-X Port-A D1 launches with poloidal aiming angles $-20^\circ$, $-10^\circ$ and $+5^\circ$. All cases use $n_e=2(1-\rho^2)\times10^{20}\ \mathrm{m}^{-3}$. The electron temperature is $T_e=3(1-\rho^2)$ keV for the $-10^\circ$ and $+5^\circ$ cases and $T_e=3.5(1-\rho^2)$ keV for the $-20^\circ$ case. The trajectories and absorbed fractions agree well for all three launch angles, as shown in Table \ref{tab:d1}.

Panels (c) of Figs. \ref{fig:w7x} and \ref{fig:d1} also compare the fully relativistic absorption model with the non-relativistic iteration-$k_i$ model of Sec. \ref{sec:absorption}. The non-relativistic model damps the waves appreciably earlier and deviates substantially from the relativistic curves, demonstrating the necessity of the relativistic treatment for these ECW cases.

\begin{table}[htbp]
\caption{Final absorbed fractions for the three W7-X Port-A D1 launches.}
\label{tab:d1}
\centering
\begin{tabular}{lll}
\hline
$\theta_{\mathrm{pol}}$ & Raytrax & BORAY-3D\\
\hline
$-20^\circ$ & $87.7\%$ & $87.3\%$\\
$-10^\circ$ & $22.0\%$ & $22.4\%$\\
$+5^\circ$ & $12.8\%$ & $11.0\%$\\
\hline
\end{tabular}
\end{table}

\begin{figure}[htbp]
\centering
\includegraphics{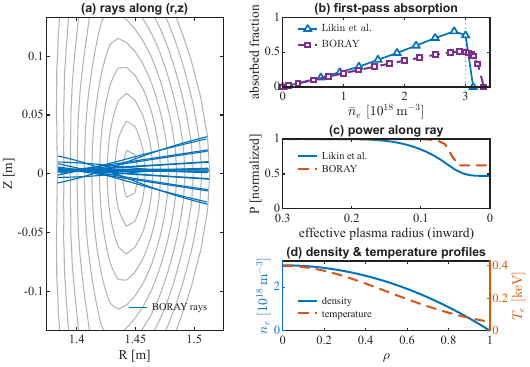}
\caption{Comparison of BORAY-3D with the HSX 28 GHz X2 heating results of Ref. \cite{Likin2003}. (a) The 18 Gaussian-beam rays at $\bar n_e=2\times10^{18}\ \mathrm{m}^{-3}$, (b) first-pass absorption versus line-averaged density, (c) averaged power from the plasma edge toward the magnetic axis and (d) density and temperature profiles.}
\label{fig:hsx}
\end{figure}

Fig. \ref{fig:hsx} compares BORAY-3D with the HSX 28 GHz X2 heating results of Likin et al.\ \cite{Likin2003}. A public HSX QHS vacuum VMEC equilibrium is scaled to $B_0=0.5$ T. The wave is launched from the low-field side and focused at the magnetic axis. The Gaussian beam is represented by 18 weighted rays with a 2 cm radius at the $e^{-2}$ power level. The profiles are $n_e=n_{e0}(1-\rho^2)$ and $T_e=T_{e0}\exp(-2\rho^2)$.

The calculated first-pass absorption increases almost linearly with density until the X-mode cutoff, in agreement with Ref. \cite{Likin2003}. At $\bar n_e=1.5\times10^{18}\ \mathrm{m}^{-3}$, BORAY-3D gives an absorbed fraction of 0.29, compared with the published value 0.40. The deposition is in the central region in both calculations, although the BORAY-3D profile is narrower. The machine-readable Biot--Savart field and individual rays of Ref. \cite{Likin2003} are unavailable. Therefore, the comparison is made for the density dependence and deposition position rather than for individual ray trajectories.

\FloatBarrier

\subsection{LHD-like ICW}
\label{sec:lhdicrf}

To show the application of BORAY-3D to ion-cyclotron waves in a stellarator, we consider the 50 MHz fast-wave case studied by Saito et al. \cite{Saito2010}. Their helical-reactor model was obtained by enlarging an LHD $R_{\rm ax}=3.75$ m configuration by a factor of 3.5. The reported rays reflect near the far-edge R cutoff, with a density-dependent $k_\parallel$ upshift and a change from tritium to electron absorption. The original numerical equilibrium and ray data are not available, so only these physical trends and characteristic values are compared here.

The vacuum field is generated from the public single-filament \texttt{lhd\_like} coil set \cite{Landreman2021SIMSOPT}. The current ratios correspond to the $R_{\rm ax}=3.75$ m and $\gamma=1.254$ configuration \cite{Watanabe2006NIFS}. The geometry is enlarged by 3.5 and the magnetic field is scaled to $B_{\rm ax}=4.92$ T. The last closed flux surface is determined by field-line tracing using the LHD criterion of Ref. \cite{Wang2012LCFS}. Since the launch point in Ref. \cite{Saito2010} is outside this flux surface, it is moved along the same direction to $\rho=0.98$. The published values $(k_{\parallel0},k_{{\rm surf},0})=(\pm0.7,\pm0.7)$ m$^{-1}$ are retained, and the surface-normal component is obtained from the inward fast-wave root of the cold-plasma dispersion relation.

\begin{figure}[htbp]
\centering
\includegraphics{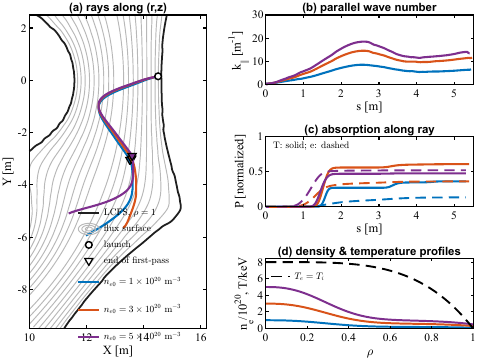}
\caption{BORAY-3D results for a 50 MHz fast wave in an LHD-like helical field. (a) Top views of the rays for $n_{e0}=1$, 3 and $5\times10^{20}$ m$^{-3}$, (b) parallel wave number, (c) first-pass tritium (solid) and electron (dashed) absorption and (d) density and temperature profiles.}
\label{fig:lhd_icrf}
\end{figure}

Equal D and T concentrations are used.  The temperature is $T_e=T_i=8(1-\rho^4)$ keV and the density is
\begin{equation}
\frac{n_e}{n_{e0}}=0.8\exp[-(\rho/0.35)^{2.5}]
+0.1(1-\rho^{6.5})+0.1,
\end{equation}
where $n_{e0}=1$, 3 and $5\times10^{20}$ m$^{-3}$. The rays plotted in Fig. \ref{fig:lhd_icrf} use $(k_{\parallel0},k_{{\rm surf},0})=(0.7,0.7)$ m$^{-1}$. Absorption is calculated with the non-relativistic hot-plasma model in Sec. \ref{sec:absorption}, from the launch to the outward $\rho=0.95$ crossing before the R cutoff.

\FloatBarrier

As shown in Fig. \ref{fig:lhd_icrf}(a), the three rays penetrate to $\rho\simeq0.20$, reflect near the far-edge R cutoff and return inward. The maximum $|k_\parallel|$ is 8.6, 14.7 and 18.6 m$^{-1}$ for increasing density, compared with approximately 8, 15 and 20 m$^{-1}$ in Ref. \cite{Saito2010}. The first-pass absorbed fraction increases from about 0.48 to nearly complete absorption over the density scan. Tritium absorption is dominant at the lowest density, while the electron share increases to about one half at the highest density, and deuterium absorption is negligible. These density dependences agree with those reported by Saito et al., although the absolute absorption cannot be compared directly because the magnetic equilibria are different.

\subsection{MUSE helicon}
\label{sec:musehelicon}

MUSE is a two-field-period permanent-magnet stellarator with three external helicon antennas \cite{Qian2023MUSE}. We use the fixed-boundary vacuum VMEC equilibrium from the public design repository. The recalculated equilibrium has $R_0=0.308$ m, $a=0.0425$ m, $B_{\rm axis}=0.147$--0.148 T and $\iota=0.181$--0.197, in agreement with the published design.

The antenna spectrum and plasma profiles have not been published, so the following calculation is an application example rather than an experimental benchmark. A hydrogen plasma with $n_e(0)=10^{20}$ m$^{-3}$, $T_e=3$ eV and $T_i=0.3$ eV is used, with
\begin{equation}
\frac{n_e(\rho)}{n_e(0)}=0.8(1-\rho^5)^5+0.2 .
\end{equation}
The frequency is 13.56 MHz. Following the $m=\pm1$ spectral component of a half-helical antenna \cite{Granetzny2025Helicon}, an effective helical length $L_h=0.10$ m gives $|k_{\parallel0}|=\pi/L_h=31.4$ m$^{-1}$. Four rays are launched inward from $\rho=0.90$ at the symmetry planes $\phi_0=0$ and $\pi/2$, with both signs of $k_\parallel$. The perpendicular wave number is given by the smaller propagating root of the cold-plasma dispersion relation. The launch conditions represent waves after coupling into the plasma.

\begin{figure}[htbp]
\centering
\includegraphics{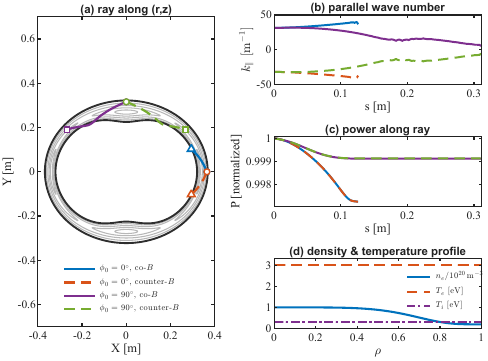}
\caption{BORAY-3D results for a 13.56 MHz helicon wave in the MUSE design equilibrium. (a) Top-view ray trajectories, (b) parallel wave number, (c) first-pass collisionless absorption and (d) assumed plasma profiles.}
\label{fig:muse_helicon}
\end{figure}
\FloatBarrier

The two rays launched at $\phi_0=0$ pass close to the magnetic axis and turn after a path length of about 0.13 m. The $\phi_0=\pi/2$ rays remain outside $\rho=0.19$ and leave the closed-flux region after about 0.31 m. The co- and counter-$B$ rays at each symmetry plane are almost identical, as required by stellarator symmetry. The different trajectories at the two launch planes show the effect of the three-dimensional magnetic geometry.

The first-pass absorbed fractions are about 0.27\% for $\phi_0=0$ and 0.09\% for $\phi_0=\pi/2$. These values include only collisionless Maxwellian damping for the assumed profiles and launch conditions. A quantitative helicon study of MUSE would require the measured plasma profiles and antenna spectrum, together with the effects of collisions, wall losses and full-wave mode coupling.

\subsection{W7-X X-mode ECE}
\label{sec:ecebench}

\begin{figure}[htbp]
\centering
\includegraphics{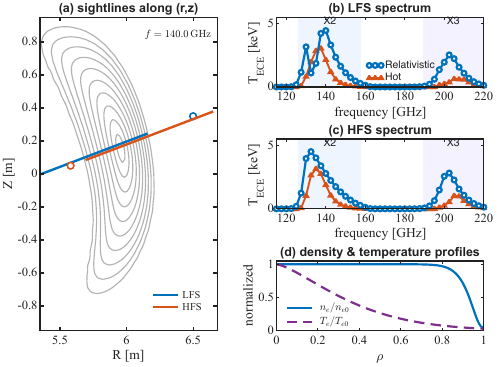}
\caption{W7-X ECE spectra calculated by BORAY-3D. (a) The $\phi=5.88^\circ$ equilibrium cross-section and 140 GHz central rays, (b) LFS spectrum, (c) HFS spectrum and (d) density and temperature profiles. Blue circles and orange triangles denote the relativistic and non-relativistic results, respectively.}
\label{fig:ece}
\end{figure}

Fig. \ref{fig:ece} shows the W7-X low-field-side (LFS) and high-field-side (HFS) ECE spectra calculated by BORAY-3D. The same public W7-X equilibrium as in Sec. \ref{sec:w7x} is used. The X-mode frequency is varied from 115 to 220 GHz in 2.5 GHz steps. The receiver positions are $(R,\phi,Z)=(6.50\ \mathrm{m},6.1^\circ,0.35\ \mathrm{m})$ for the LFS and $(5.58\ \mathrm{m},5.3^\circ,0.05\ \mathrm{m})$ for the HFS. Each spectrum is averaged over a central ray and four rays at the measured $1.2^\circ$ beam divergence. The density is nearly flat with $n_e(0)=2\times10^{19}\ \mathrm{m}^{-3}$, and $T_e/T_e(0)=\exp(-3.5\rho^{1.4})$ with $T_e(0)=5$ keV. The receiver geometry is taken from Refs. \cite{Marushchenko2007,Marushchenko2006}, and the temperature profile is fitted to the W7-X measurements in Ref. \cite{Zanini2020}.

Both the X2 and X3 emission bands are obtained. The X2 and X3 peak radiative temperatures are 4.5 and 2.5 keV for the LFS and 4.5 and 2.8 keV for the HFS, respectively. The X3 band has an optical depth of order unity. In comparison, the non-relativistic absorption model underestimates the X3 optical depth by about one order of magnitude.

\section{Summary and discussion}
\label{sec:summary}

A new three-dimensional RF wave ray tracing code BORAY-3D has been developed and validated for general configuration of magnetized plasma in this work. The code exhibits excellent agreements with GENRAY code for tokamak ripple cases of LHW, Raytrax code and TRAVIS code for W7-X stellarator case of ECW, as well as public simulation results on ICRF of LHD case and ECE case of W7-X. The main advantage of BORAY-3D is the combination of the following three capabilities that are not considered simultaneously in previous studies:

First, it provides broad-frequency-range ray modeling within a single theoretical and numerical framework, from ICW, helicon and LHW to ECW. The wave frequency of application covers the range of 13.56 MHz--220 GHz.

Second, it applies cylindrical coordinates to implement the physics model equations, which are applicable to arbitrary two-dimensional axisymmetric and fully three-dimensional magnetic-plasma configurations without the constraint of flux coordinates and faithfully incorporate both closed and open field-line regions.

Third, a fully relativistic Maxwellian EC absorption and reciprocal ECE radiative transfer model is formulated, which resolves the limitation of the original BORAY code \cite{Xie2022}.

Considering the limitation of geometric optical approximation, the wave reflection at the plasma edge cannot be addressed in a physical manner by ray tracing codes, as well as other effects such as diffraction. Full-wave modeling is generally required for quantitative validation in such cases. As in BORAY, the density and temperature need not be magnetic-flux functions, and no special treatment is required for open field-line regions. We will focus on collisional damping, wall reflection, current drive and non-Maxwellian distributions. Moreover, the present scheme is appropriate for the waves close to the cold plasma dispersion relation, and we will also improve the cold plasma model to kinetic dispersion relation for electron and ion Bernstein waves, which we demonstrate at Ref. \cite{xie2021plasma}, in future work.

\ack{The authors thank Jian Bao for his many helpful suggestions, and Wanying Yu for helpful discussions on the absorption model and the algorithm for calculating species absorption ratio.}

\section*{Appendix}
\appendix
\section{Auxiliary tensors for relativistic EC absorption}
\label{app:relativistic}

This appendix gives the Krivenski--Orefice response tensor used in the relativistic EC absorption model of Sec. \ref{sec:ece} and the non-relativistic R\"onnmark hot-plasma tensor.
Let $\mathcal{F}_{q}(s)$ denote the weakly relativistic Shkarofsky function of Ref. \cite{Krivenski1983}. For non-negative harmonic index $s$ and finite-Larmor-radius order $\ell$, define $\lambda=\mu^{-1}(N_{\perp}\omega/\Omega_e)^2$ and $\eta=N_{\perp}\omega/\Omega_e$, and introduce
\begin{equation}
\begin{aligned}
a_{s\ell}&=
\frac{(-1)^\ell\Gamma(s+\ell+\tfrac12)}
{2^s\Gamma(\ell+1)
\Gamma(s+\tfrac{\ell}{2}+1)
\Gamma(s+\tfrac{\ell}{2}+\tfrac12)},\\
b_{s\ell}&=
\left[(s+\ell)^2-
\frac{\ell(\ell+2s)}{2s+2\ell-1}\right]a_{s\ell},
\end{aligned}
\label{eq:koab}
\end{equation}
where $\Gamma$ is the gamma function. Let $\nu=s+\ell+3/2$ and define
\begin{equation}
\begin{aligned}
Q_{s\ell}^{(0)}&=\mathcal{F}_{\nu}(s),\\
Q_{s\ell}^{(1)}&=N_{\parallel}
\left[\mathcal{F}_{\nu}(s)-\mathcal{F}_{\nu+1}(s)\right],\\
Q_{s\ell}^{(2)}&=\frac{\mathcal{F}_{\nu+1}(s)}{\mu}
+N_{\parallel}^{2}
\left[\mathcal{F}_{\nu+2}(s)+\mathcal{F}_{\nu}(s)
-2\mathcal{F}_{\nu+1}(s)\right].
\end{aligned}
\label{eq:koq}
\end{equation}
For $s>0$, set $Q_{s\ell}^{(h,\pm)}
=Q_{s\ell}^{(h)}\pm Q_{-s,\ell}^{(h)}$. For $s=0$, set
$Q_{0\ell}^{(h,+)}=Q_{0\ell}^{(h)}$ and
$Q_{0\ell}^{(h,-)}=0$. The six independent elements of the Krivenski--Orefice response matrix $\boldsymbol{G}$ are
\begin{equation}
\begin{aligned}
G_{11}&=\sum_{s,\ell}s^2a_{s\ell}\lambda^{s+\ell-1}
Q_{s\ell}^{(0,+)},\\
G_{12}&=-G_{21}=
i\sum_{s,\ell}s(s+\ell)a_{s\ell}\lambda^{s+\ell-1}
Q_{s\ell}^{(0,-)},\\
G_{22}&=\sum_{s,\ell}b_{s\ell}\lambda^{s+\ell-1}
Q_{s\ell}^{(0,+)},\\
G_{13}&=G_{31}=
\eta\sum_{s,\ell}s\,a_{s\ell}
\lambda^{s+\ell-1}Q_{s\ell}^{(1,-)},\\
G_{23}&=-G_{32}=
-i\eta\sum_{s,\ell}(s+\ell)a_{s\ell}
\lambda^{s+\ell-1}Q_{s\ell}^{(1,+)},\\
G_{33}&=\mu\lambda\sum_{s,\ell}a_{s\ell}
\lambda^{s+\ell-1}Q_{s\ell}^{(2,+)}.
\end{aligned}
\label{eq:komatrix}
\end{equation}
The sums run over non-negative $s$ and $\ell$. Terms that contain a vanishing power of $\lambda$ are understood in their continuous $\lambda\rightarrow0$ limit. The dielectric and dispersion matrices are
$\boldsymbol{\epsilon}_{\mathrm{KO}}=\boldsymbol{I}-\mu(\omega_{pe}^{2}/\omega^2)\boldsymbol{G}$ and $\boldsymbol{\mathcal{D}}_{\mathrm{KO}}=\boldsymbol{\epsilon}_{\mathrm{KO}}-N^2\boldsymbol{I}+\boldsymbol{N}\boldsymbol{N}^{\mathsf T}$, respectively. Here $\boldsymbol{I}$ is the $3\times3$ identity matrix and $N^2=N_{\perp}^2+N_{\parallel}^2$.

For the R\"onnmark tensor, $s$ denotes the plasma species. Define $v_{ts}=\sqrt{2T_s/m_s}$, $\omega_{cs}=q_sB/m_s$, $\rho_{cs}=v_{ts}/(\sqrt{2}\omega_{cs})$, $a_s=k_{\perp}\rho_{cs}$, $\lambda_s=a_s^2$, $\omega_{ps}^2=n_sq_s^2/(\epsilon_0m_s)$ and $x_{sj}=(\omega-k_zv_{ts}c_j)/\omega_{cs}$, where $k_z=k_{\parallel}$. Here $T_s$, $m_s$, $q_s$ and $n_s$ are the temperature, mass, charge and density, respectively. The coefficients $b_j$ and $c_j$ are defined by the Pad\'e representation $Z(\zeta)\simeq\sum_j b_j/(\zeta-c_j)$. With $\Gamma_n(\lambda)=I_n(\lambda)e^{-\lambda}$, the R\"onnmark functions of Ref. \cite{Yu2026Absorption} are
\begin{equation}
\begin{aligned}
R(x,\lambda)
&\equiv\sum_{n=-\infty}^{\infty}
\frac{n^2\Gamma_n(\lambda)}{\lambda(x-n)}
=-\frac{x}{\lambda}
+x^2\sum_{n=-\infty}^{\infty}
\frac{\Gamma_n(\lambda)}{\lambda(x-n)},\\
R'(x,\lambda)
&\equiv\frac{\partial[\lambda R(x,\lambda)]}{\partial\lambda}
=\sum_{n=-\infty}^{\infty}
\frac{n^2\Gamma'_n(\lambda)}{x-n}
=x^2\sum_{n=-\infty}^{\infty}
\frac{\Gamma'_n(\lambda)}{x-n}.
\end{aligned}
\label{eq:ronnmarkfunctions}
\end{equation}
With $R_{sj}=R(x_{sj},\lambda_s)$ and $R'_{sj}=R'(x_{sj},\lambda_s)$, the R\"onnmark response tensor is
\begin{equation}
\boldsymbol{\epsilon}_{\mathrm R}
=\sum_s\frac{\omega_{ps}^{2}}{\omega\omega_{cs}}
\sum_j b_j
\begin{pmatrix}
R_{sj} & \dfrac{i}{x_{sj}}R'_{sj} &
\dfrac{\sqrt{2}a_sc_j}{x_{sj}}R_{sj}\\
-\dfrac{i}{x_{sj}}R'_{sj} &
R_{sj}-\dfrac{2\lambda_s}{x_{sj}^{2}}R'_{sj} &
-\dfrac{i\sqrt{2}a_sc_j}{x_{sj}^{2}}R'_{sj}\\
\dfrac{\sqrt{2}a_sc_j}{x_{sj}}R_{sj} &
\dfrac{i\sqrt{2}a_sc_j}{x_{sj}^{2}}R'_{sj} &
\dfrac{2c_j^2}{x_{sj}^{2}}(x_{sj}+\lambda_sR_{sj})
\end{pmatrix}.
\label{eq:ronnmarktensor}
\end{equation}

Let $\boldsymbol{K}_{\mathrm R}=\boldsymbol{I}+\boldsymbol{\epsilon}_{\mathrm R}$ and $\boldsymbol{K}_{\mathrm R}^{H}
=(\boldsymbol{K}_{\mathrm R}+\boldsymbol{K}_{\mathrm R}^{\dagger})/2$.
Holding the local unit polarization fixed, define the scalar Hamiltonian
\begin{equation}
\mathcal{H}_{\mathrm R}=
\widehat{\boldsymbol{e}}^{\dagger}
\left[
N^2\boldsymbol{I}-\boldsymbol{N}\boldsymbol{N}^{\mathsf T}
-\boldsymbol{K}_{\mathrm R}^{H}
\right]\widehat{\boldsymbol{e}} .
\label{eq:fluxhamiltonian}
\end{equation}
The dimensionless power-flux vector is
\begin{align}
F_j
&=\frac12\frac{\partial\mathcal{H}_{\mathrm R}}{\partial N_j}\nonumber\\
&=N_j-\operatorname{Re}
\left[(\boldsymbol{N}\cdot\widehat{\boldsymbol{e}})^*e_j\right]
-\frac12\operatorname{Re}\left[
\widehat{\boldsymbol{e}}^{\dagger}
\frac{\partial\boldsymbol{K}_{\mathrm R}^{H}}{\partial N_j}
\widehat{\boldsymbol{e}}\right],
\qquad j=x,y,z ,
\label{eq:hotflux}
\end{align}
where $N_x=N_{\perp}$, $N_y=0$, $N_z=N_{\parallel}$ and
$e_j$ is the corresponding component of
$\widehat{\boldsymbol{e}}$.

\bibliographystyle{iopart-num}
\bibliography{boray3d}

\end{document}